\documentclass[twocolumn,appendixfloats,tighten]{aastex6}
\usepackage{newtxtext,newtxmath}
\usepackage[T1]{fontenc}
\usepackage{ae,aecompl}

\usepackage{amsmath}	
\usepackage{amssymb}	

\usepackage{ctable}
\usepackage{url}
\usepackage{xspace}
\usepackage[normalem]{ulem}
\usepackage{hyperref}
\usepackage[all]{hypcap}
\usepackage{graphicx}
\usepackage{color}
\def\msun{{\rm\,M_\odot}}

\def\msun{{\rm\,M_\odot}} 
\def\lsun{{\rm\,L_\odot}}

\newcommand{\kms}{\, {\rm km\, s}^{-1}}

\newcommand{\be}{\begin{equation}}
\newcommand{\ee}{\end{equation}}

\newcommand{\mstar}{M_{\star}}
\newcommand{\lstar}{L_{\star}}

\newcommand{\tmerge}{t_{\rm merge}}
\newcommand{\vcm}{\rm{v_{CM}}}

\def\h2{${\rm\,H_2}$}

\newcommand{\rsun}{R_{\odot}}

\newcommand{\rprocess}{\emph{r}-process }
\newcommand{\vesc}{\rm v_{esc} }
\def\gsim{ \lower .75ex \hbox{$\sim$} \llap{\raise .27ex \hbox{$>$}} }
\def\lsim{ \lower .75ex \hbox{$\sim$} \llap{\raise .27ex \hbox{$<$}} }

\usepackage{color}

\begin{document}

\title{$r$-process enrichment of the Ultra-Faint Dwarf Galaxies by Fast Merging Double Neutron Stars}
\author{Mohammadtaher Safarzadeh\altaffilmark{1}, Enrico Ramirez-Ruiz\altaffilmark{2,3}, Jeff. J. Andrews\altaffilmark{3,4,5}, Phillip Macias\altaffilmark{2}, Tassos Fragos \altaffilmark{6}, Evan Scannapieco \altaffilmark{1}}

\altaffiltext{1}{ School of Earth and Space Exploration, Arizona State University, \href{mailto:mts@asu.edu}{mts@asu.edu}}
\altaffiltext{2}{Department of Astronomy and Astrophysics, University of California, Santa Cruz, CA 95064, USA}
\altaffiltext{3}{Niels Bohr Institute, Blegdamsvej 17, 2100 K\o benhavn \O, Denmark}
\altaffiltext{4}{Foundation for Research and Technology-Hellas, 100 Nikolaou Plastira St., 71110 Heraklion, Crete, Greece}
\altaffiltext{5}{Physics Department \& Institute of Theoretical \& Computational Physics, University of Crete, 71003 Heraklion, Crete, Greece}
\altaffiltext{6}{Geneva Observatory, University of Geneva, Chemin des Maillettes 51, 1290 Sauverny, Switzerland }
\begin{abstract}
The recent aLIGO/aVirgo discovery of gravitational waves from the neutron star merger (NSM) GW170817 and the follow up kilonova observations have shown that NSMs produce copious amount of \rprocess  material.  However, it is difficult to reconcile the large natal kicks and long average merging times of Double Neutron Stars (DNSs), with the levels of \rprocess enrichment seen in ultra-faint dwarf (UFD) galaxies such as Reticulum II and Tucana III.  Assuming that such dwarf systems have lost a significant fraction of their stellar mass through tidal stripping, we conclude that contrary to most current models, it is the DNSs with rather large natal kicks but very short merging timescales that can enrich UFD-type galaxies. These binaries are either on highly eccentric orbits or form with very short separations due to an additional mass-transfer between the first-born neutron star and a naked helium star, progenitor of the second-born neutron star.
These DNSs are born with a frequency that agrees with the statistics of the \rprocess UFDs, and merge well within the virial radius of their host halos, therefore contributing significantly to their \rprocess enrichment.
\end{abstract}

\section{Introduction}

The recent aLIGO/aVirgo discovery of a neutron star merger GW170817 \citep{Collaboration:2017kt} and the subsequent kilonova observed across the entire electromagnetic spectrum \citep{Abbott:2017it,Coulter:2017ei,Drout:2017gb,Margutti:2018ek} has undoubtedly shown that neutron star mergers (NSMs) produce 
\rprocess elements in copious amounts \citep{Kasen:2017kk}.
Prior to this discovery, two ultra-faint dwarf (UFD) galaxies, Reticulum II \citep{Ji:2016ja}, and Tucana III \citep{Hansen:2017ho}, had been observed to be enriched in \rprocess elements, the statistics of which could be explained with a single rare event such as a neutron star merger.
These galaxies were discovered by the Dark Energy Survey \citep{Koposov:2015cw,DrlicaWagner:2015gb}, and are the only two UFDs known to show substantial \rprocess enrichment, despite observational efforts to identify others \citep{Ji:2018wr}.

UFDs \citep{Brown:2012jo,Frebel:2012ja,Vargas:2013ei} are satellites of the Milky Way that are highly dark matter dominated  \citep{Simon:2007ee} with total luminosities of $\lstar\approx10^3-10^5\lsun$, and they were discovered in deep wide-area sky surveys \citep{Koposov:2015cw,Koposov:2015ep,Bechtol:2015bd}.
Based on color-magnitude diagram analysis, \citet{Brown:2014jn} finds that nearly 3/4 of the entire stellar mass content of such galaxies is formed by $z\approx10$ and $\approx80$\% of the stellar mass content is already formed by $z\approx6$. 

It has been challenging to explain the observed \rprocess enrichment of UFDs with NSM events due to two facts: (i) UFD halo progenitors have shallow potential wells, corresponding to halo masses, $M_{\rm h}$, between $10^8$ and $10^9\msun,$ depending on the assumed dark matter density profile \citep{Simon:2007ee,Bovill:2009hg,BlandHawthorn:2015ke}. The halo mass probability density function (PDF) of UFD progenitors has a maximum of $M_{\rm h}\approx10^9 \msun$ at redshifts around $z\approx10$ \citep{Safarzadeh:2018fi}. Such halos have escape velocities $\vesc\approx25\kms$, with the majority of halos of plausible UFD progenitors having $\vesc$ less than 10 $\kms$.
(ii) NSs are born with kicks that may be as large as hundreds of $\kms$ \citep{Hobbs:2005be,Janka:2017}. Although there is evidence that at least some NSs receive lower kicks at birth \citep[e.g.,][]{Pfahl:2002,van_den_Heuvel:2007,Verbunt:2017}, five of the six DNSs with 3D velocities provided by \citet{Tauris:2017cf} have peculiar velocities larger than 25 $\kms$, which suggests they would have escaped their host halos and therefore would not contribute to their \rprocess enrichment if they were born in UFD-type galaxies.

To have a closer look at the issue, one must take into account the history of star formation of UFD-type systems, which ceases during reionization \citep{Bovill:2009hg,Bovill:2011bk,Brown:2014jn,Weisz:2014cp}. This implies that the NSM events should have taken place in less than about 1 Gyr since the start of star formation. Combined with the shallow potential well of UFD host halos, if an NSM is to be considered as a plausible source of \rprocess enrichment, a binary with a merging time of less than 1 Gyr, and an escape velocity less than 25 $\kms$ needs to be formed frequently in the early universe, and different theoretical studies disagree on whether such binaries could have formed in UFDs \citep{Beniamini:2016kwa,Bramante:2016kp}.

Separately, through cosmological zoom simulations of UFDs, \citet{Safarzadeh:2017dq} demonstrated that a single NSM event can account for the observed \rprocess
enrichment in Reticulum II, with modest assumptions regarding the europium yield in the dynamical ejecta of the NSM. However, the authors had to make the NSM event take place in less than about 10 Myr after the onset of star formation in the UFDs so that the subsequent
generation of stars can inherit the \rprocess ejected into the interstellar medium (ISM) before the total stellar mass of the system reaches $\mstar\approx10^4\msun$. The simulations were stopped when
the total stellar mass reached the level required to mimic the suppression of star formation by reionization.  The NSM events modeled were  effectively from fast merging DNSs with very low natal kicks.
Moreover, it has been pointed out that small natal kicks can have a significant impact on \rprocess enrichment of UFDs \citep{Safarzadeh:2017dq}, as well as the MW galaxy \citep{Behroozi:2014bp,Safarzadeh:2017dw}.

In this study, we consider different pathways through which UFDs could have been enriched in \rprocess material.  We combine our understanding of star formation and halo assembly of such systems with 
publicly available population synthesis models of binary compact object formation \citep{Dominik:2012cw}.

The structure of this work is as follows.
In \S2 we describe the population synthesis model that we have analyzed in our study.
In \S3 we present our results showing where in the parameter space the DNS candidates are born in such models that could enrich UFD-like systems with \rprocess material.
In \S4 we discuss in detail how fast-merging DNSs are formed.
In \S5 we compare the birth rate of the candidate DNSs to observations of the \rprocess enriched UFDs, and 
in \S6 we summarize our results and discuss the future work needed to improve our understanding of \rprocess enrichment at high redshifts.  

\section{Method}
To model the formation of double neutron stars (DNSs), we use published results from the StarTrack \citep{Belczynski:2002gi,Belczynski:2006dq,Belczynski:2008kt} population synthesis code. These simulations are described in detail in \citet{Dominik:2012cw} and include three major improvements over previous StarTrack versions with regards to stellar winds \citep{Belczynski:2010iw}, common envelope formulation, and 
compact object formation \citep{Fryer:2012jk}. We have analyzed the intermediate data from these simulations that are publicly available\footnote{The public data from the Synthetic Universe Project is available at: www.syntheticuniverse.org}. We briefly summarize the method in this section.

Each of the models we analyze has $2\times10^6$ binaries, initialized by four different parameters: (i) the primary star's mass $M_1$, 
(ii) the mass ratio $q=M_2/M_1$ with $M_2$ being the mass of the secondary star, (iii) the semi-major axis $a$ of the binary, and (iv) the eccentricity $e$. The masses of the primary are drawn from Kroupa initial mass function (IMF)
from 5 to 150 $\msun$. The mass ratio is assumed to have a flat distribution between $q=0-1$ with the minimum mass of the secondary considered to be 3 $\msun$. The distribution of initial binary separation is assumed to be flat in log(a), and the eccentricity is drawn from thermal equilibrium distribution $\Xi(e)=2e$, with $e$ ranging from $0$ to $1$. 

\subsection{Model Variations}
A number of parameters are varied among the models, which are meant to parametrize the uncertainties in our understanding of binary neutron star formation. 
The first parameter that is varied deals with the behavior of binaries when a star enters into a common envelope (CE) with a Hertzsprung gap (HG) donor star; during a CE, a star enters the envelope of its companion, exchanging orbital energy to unbind the donor's envelope. For giant stars, with a clear core-envelope boundary, the end result of this process (so long as there is enough orbital energy available to keep the system from merging) is a closely bound binary comprised of the accretor star and the giant star's core. However, HG stars lack well-defined cores, and studies are inconclusive as to whether binaries entering into a CE during this phase can survive without merging \citep{Deloye:2010ey}. We note that allowing HG stars to survive a CE significantly increases the merger rates of double black hole binaries to a level that exceeds current LIGO constraints \citep{Belczynski:2007}, although it matches NSM rate estimates \citep{Chruslinska:2018bv}. 

Two different sub-models, A and B have been analyzed which treat differently CE event with HG donor stars.
Submodel A treats HG stars such that a core could be 
distinguished from an envelope in their evolutionary phase, hence a successful CE ejection is possible, while submodel B assumes any system entering into a CE with a HG donor will merge. As a second parameter, the kick velocity received by a NS at birth is varied. Our standard model adopts natal kicks randomly drawn from a Maxwellian distribution with $\sigma=265\kms$, based on the observed velocities of single Galactic pulsars \citep{Hobbs:2005be}, and we explore the models that adopt $\sigma=135\kms$. We note that to match merger rate estimates from LIGO, recent binary population synthesis results suggest that DNSs are likely to be born with kick velocities lower than the low kick models explored in this work \citep{Giacobbo:2018}. Finally, we test models with two different metallicities: $Z=Z_{\odot}$ and $Z=0.1Z_{\odot}$.

Our naming convention divides into submodel A and B which correspond to whether HG stars survive a CE or are assumed to merge, respectively. We further subdivide our models based on metallicity (either solar or 0.1 solar). We name each model as (A,B)Z(02,002)(H,L) where the first letter denotes submodels A or B, 02 and 002 correspond to solar and tenth of solar metallicity models, and the last letters (H,L) determine whether the natal kicks to the compact objects are drawn from a Maxwellian distribution with $\sigma=(265,135)\kms$ respectively. 
In summary, we test two different models with four possible variations each, producing eight separate cases.

\subsection{Definition of Candidate DNSs}

Unlike previous studies that solely focused on the natal kicks of the DNSs to determine whether they can enrich UFDs with \rprocess elements \citep{Beniamini:2016kwa,Bramante:2016kp}, here
we take a different approach: a DNS with a large systemic velocity that nevertheless merges well within the virial radius of the host halo will contribute to the \rprocess enrichment. Therefore, all the DNSs that satisfy $t_{\rm merge} \times \rm{v_{CM}} <\epsilon \times r_{\rm vir}$ are considered candidates for the enrichment of UFDs.
We set $\epsilon=0.1,$ noting that the half-light radius of a galaxy is correlated with its halo virial radius as $R_e=0.015 r_{\rm vir}$ \citep{Kravtsov:2013cy} and therefore our choice for $\epsilon$ is large enough to encompass the extended ISM of the galaxy. However, we show below that our results are rather insensitive to moderate variations in $\epsilon$. Here $ t_{\rm merge}$ is the merging time since the formation of the DNS, $\rm{v_{CM}}$ is the center of mass velocity of the binary after the formation of the DNS, and $r_{\rm vir}$ is the 
virial radius of the UFD host halo, which is a function of both the halo's mass and its redshift.

For our analysis, we consider a UFD host halo of $10^9\msun$ at $z=6$ which has a virial radius of $\approx 4.6$ kpc and escape velocity of $\rm v_{esc}\approx44\kms$. This halo mass corresponds to the maximum mass that a UFD progenitor can have at such redshifts based on various abundance matching techniques \citep{Safarzadeh:2018fi}. We note that although the definition of candidate DNSs above ignores the gravitational potential of UFDs, DNSs with short merger times typically have systemic velocities in excess of $10^2\kms$, much larger than the escape velocities of UFDs. 
If anything, our estimates should be considered conservative, in the sense that a contribution from the host galaxies potential would result in a deceleration of newly formed DNS.

While we present our results based on a virial radius cut of UFD host halos at high redshifts, one might wonder how sensitive our rates are to the assumptions regarding the halo mass of these systems. Moreover, if the DNSs merge far from the star-forming region of the galaxies, whether the released \rprocess material is recycled into the newly formed stars at high concentration to make \rprocess metal-poor stars remains to be explored with hydro simulations. \citet{Safarzadeh:2017dq} showed that the location of the NSM with respect to the star-forming region could severely affect the level of \rprocess enhancement that would be observed in the stars. 
We return to this point in the following section, where we show the distances traveled by our DNS candidates before they merge are typically less than 100 parsecs, and therefore our conclusions are robust with respect to different assumptions with regards to the virial radius of such halos at high redshifts.

\section{Results}
 We summarize our application of DNS mergers produced by the population synthesis models of \citet{Dominik:2012cw} to UFDs in Table 1. The first column presents different models studied in this work. The second column indicates the number of candidate binary NSs, which are defined as those that are born and merge well within the virial radius of a halo with mass $10^9\msun$ ($r_{\rm vir}\approx 4.6$ kpc) at $z=6$. The numbers in the parenthesis show the number of 
candidates that merge within the virial radius of a halo with mass $10^8\msun$ ($r_{\rm vir}\approx 1.3$ kpc) at $z=10$. The third column indicates the total number of surviving DNSs in each model.
 
 \begin{deluxetable}{cccccc} 
\tabletypesize{\footnotesize} 
\tablecolumns{6} 
\setlength{\tabcolsep}{3mm}
\tablewidth{0pt}
\tablecaption{Formation rate of the DNSs in different models. \label{t.sim_params}} 
\tablehead{
\colhead{  Model } & \colhead{\# of candidate DNSs} & \colhead{\# of all DNSs} & \colhead{Fraction} }
\startdata 
AZ002H			&1041(920)  		& $5934 $ &	18\%	\\
AZ002L			&1205(1139)		& $9918 $ 	&12\%	\\
AZ02H			&678(561) 		& $9713 $ 	&7\%	\\
AZ02L			&1255(1158)  		& $15575 $ &8\%	\\
BZ002H			&52(35)		& $4241 $	&1\%	\\
BZ002L			&17(12)		& $7620 $	&0.2\%	\\	
BZ02H			&112(65) 		& $6061 $	&2\%	\\	
BZ02L			&28(18)		& $10280 $	&0.3\%	\\	
\enddata
\vspace{-0.cm}
\tablecomments{ The number of double neutron stars formed out of $2\times10^6$ binaries simulated in each model. 
The models are described in section 2.1. The second column shows the number of candidate double neutron stars, 
which are defined as those merging well within the virial radius of a halo with mass $10^9\msun$ ($r_{\rm vir}\approx 4.6$ kpc) at $z=6$. The numbers in the parenthesis show the number of 
candidates that merge within the virial radius of a halo with mass $10^8\msun$ ($r_{\rm vir}\approx 1.3$ kpc) at $z=10$. The third column indicates the total number of surviving DNSs, and the forth column shows the percentage of all the DNSs that are considered to be candidates.} 
\end{deluxetable}

For example, considering models with high kick velocities and metallicity $Z=0.1 Z_{\odot}$, out of all the $2\times10^6$ binaries, 5934 DNSs survive in submodel A and 4241 in submodel B. 
The corresponding number of DNSs that are identified as candidate DNSs are 1116 ($\approx 18.5\%)$ in submodel A, and only 92 candidates ($\approx 1.8\%)$ in submodel B. We note that If we assume Poisson noise for the numbers here, the noise would be proportional to Sqrt(N), even for the cases where only 0.1\% of all the DNSs become the candidate DNSs, the S/N is significant since the sample size is about a thousand. Therefore, although a million is not a large sample for a pop-synthesis analysis, we believe the S/N is already large enough to be secure from large statistical fluctuations.

Given the specific IMF adopted in the studies of \citet{Dominik:2012cw}, modeling $2\times10^6$ binaries corresponds to a total stellar mass of $\approx 3\times10^8\msun$. This conversion implies that one DNS is formed in (AZ002H,BZ002H)  per ($5\times10^4$, $7\times10^4$) solar mass of stars. 
Subsequently, the candidate DNSs have a birth rate of one per $2.8\times10^5\msun$ and $5.7\times10^6\msun$, respectively. 

\begin{figure*}
\setlength{\tabcolsep}{1mm}
\hspace{0cm}
\includegraphics[width=1\textwidth]{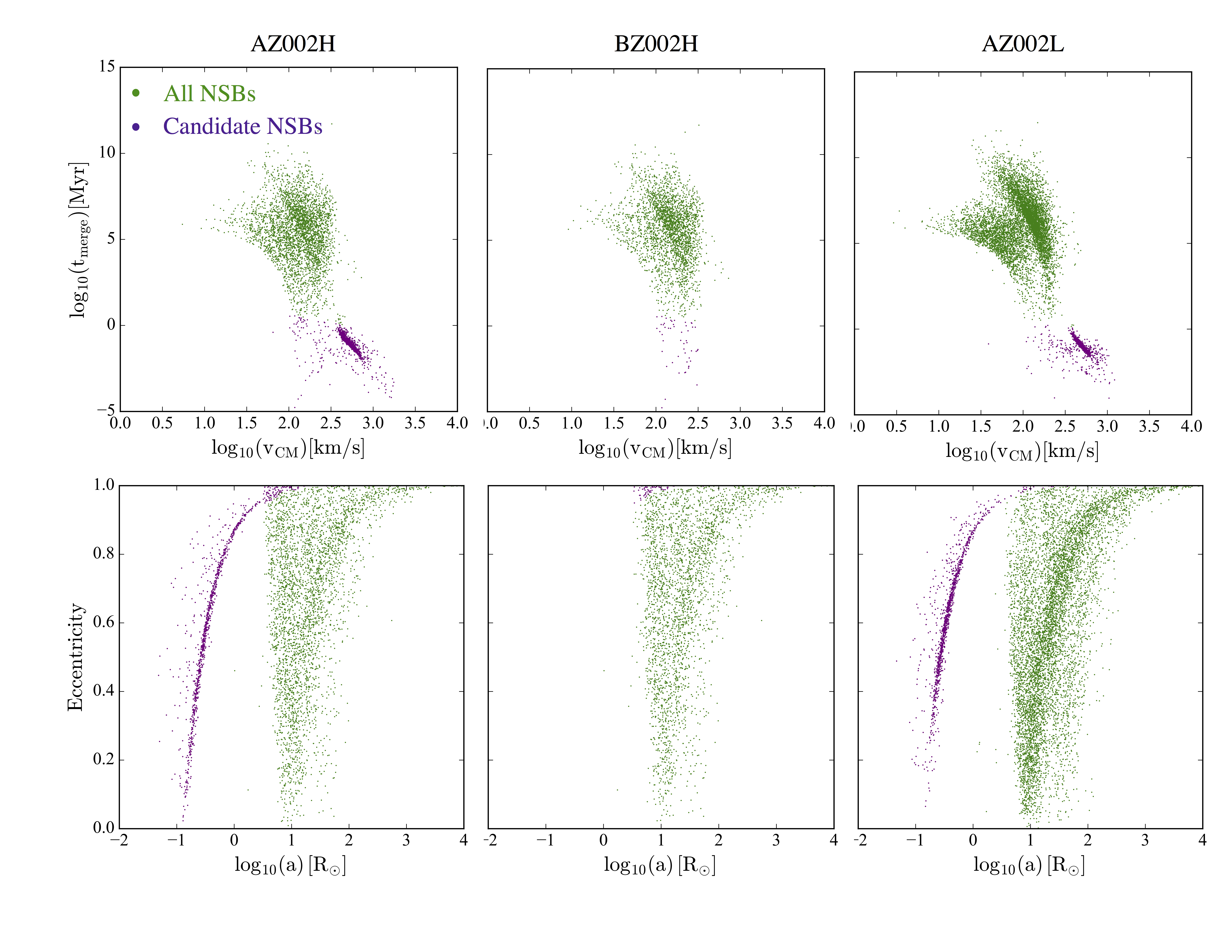}
\caption{The distribution of DNSs properties just after the formation of the second NS in the $\tmerge-\vcm$ plane (top row), and in the semi-major axis vs. eccentricity plane (bottom row). 
The green points are all the DNSs in submodel A and the purple points are the candidates DNSs that are born in $10^9\msun$ halo at $z=6$ and merge within its virial
radius. Left column corresponds to submodel A, the middle column to submodel B, and the right column to a variation of the model with low natal kicks imparted to the neutron stars upon birth. Note that we are only showing the DNSs that merge in less than $10^{13}$ Myr, and the models are at 0.1 $Z_{\odot}$.
}
\label{f:all_models}
\end{figure*}

Figure \ref{f:all_models} shows the distribution of the candidate DNSs in purple dots (and all the DNSs in a given model in green dots) in $\tmerge-\vcm$ plane (top row of panels), and in semi-major axis vs.\ eccentricity plane (bottom row of panels). The left column corresponds to model AZ002H, the middle column to BZ002H, and the right column to AZ002L. 

The various clusters in these panels are formed from different DNS formation channels, and the various choices of model parameters affects both the characteristics of DNSs formed through these channels and their relative ratios \citep{Andrews:2015bq}. The two clumps of predominantly green points (seen clearly in the top panels) are differentiated by whether the DNSs went through (stable) Case BB mass transfer during formation. Case BB mass transfer refers to when a NS accretes He from the He HG star \citep{Delgado:1981vo}.
Systems with wide enough separations, or those that after the first CE phase, have a helium star with mass larger than $3-4\msun$, do not go through case BB mass transfer, and therefore form DNSs with separations too large to merge within a Hubble time. Alternatively, systems that go through stable Case BB mass transfer tend to form shorter period systems, a fraction of which merge quickly enough to be LIGO sources. A third formation channel exists (the dense clump of purple points in the top panels) for sub-model A only, which allows for certain systems to form through unstable Case BB mass transfer. We further describe this channel -- including why it only exists for sub-model A -- below in Section \ref{ss.BB}.

Figure \ref{f:all_models} shows that there are two channels to make the candidate DNSs: (i) binaries with rather small separations in highly eccentric orbits and (ii) binaries with small separation without bias in eccentricity.
The first channel contributes to the binary formation in both submodels, while the second channel produces the majority of binaries in submodel A, but does not contribute to the formation of DNSs in submodel B. 
We discuss these channels in more detail in Section \ref{ss.channels}. 

\begin{figure}
\setlength{\tabcolsep}{1mm}
\hspace{0cm}
\includegraphics[width=\columnwidth]{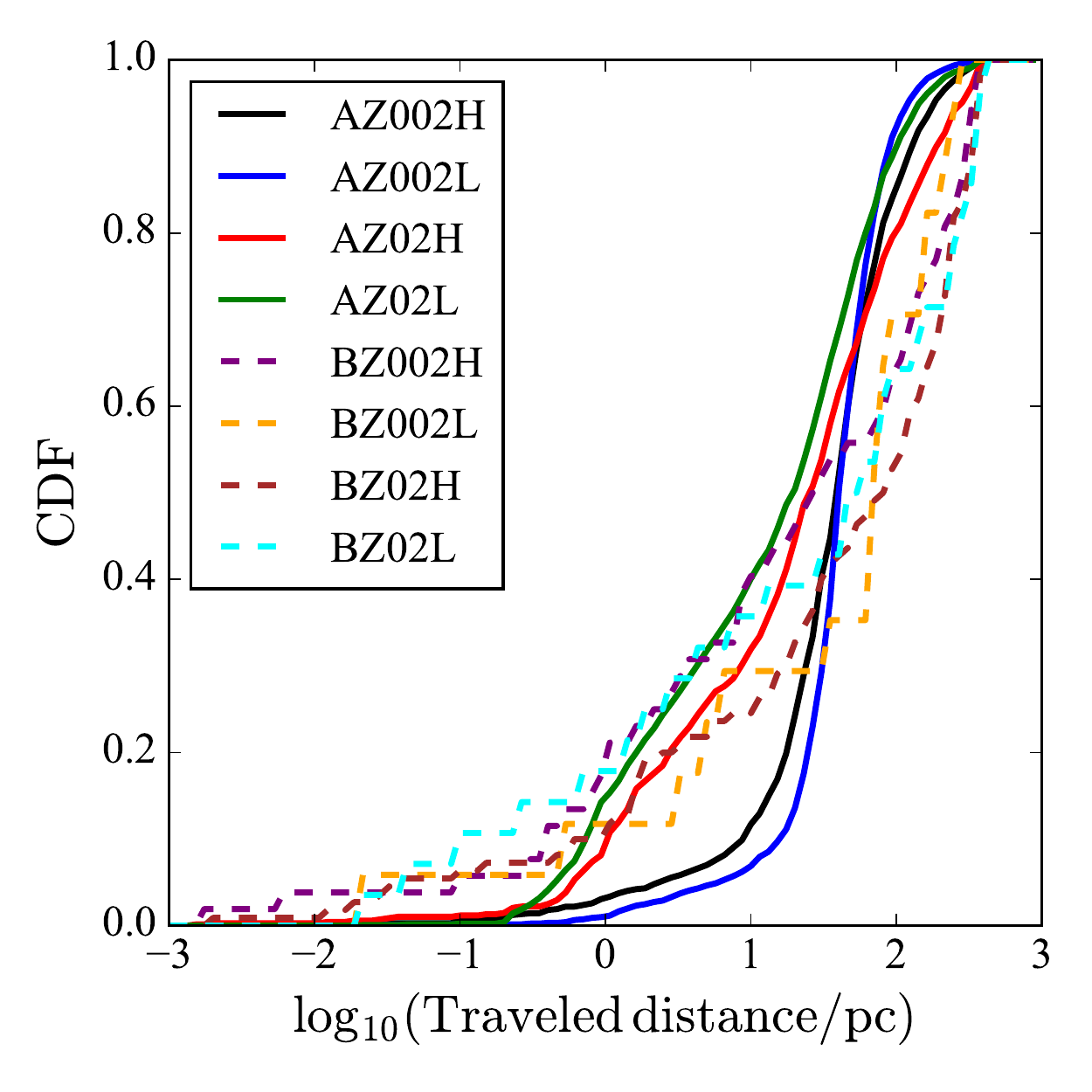}
\caption{The cumulative distribution function of the distances traveled by the DNS candidates before they merge. Regardless of the model, most of the candidates merge before they travel more than 100 parsecs.
This is well inside the potential well of the host halo and therefore a significant fraction of that the ejected \rprocess material is expected to be inherited by the subsequent star formation.
}
\label{f.distance}
\end{figure}

Figure \ref{f.distance} shows the cumulative distribution function of the distance traveled by the candidate DNSs before they merge. As can be seen, about 90\% of our candidates merge
within the central hundred parsecs of the galaxy; even though we define our candidate DNSs as those that merge within the virial radius, most candidate DNSs merge much deeper in the potential well of the galaxy. Depending on how close the NSM event takes place to the star-forming region of a UFD, different classes of \rprocess enhanced metal-poor (MP) stars \citep{Beers:2005kn}, either MP-$r$I (with 0.3<[Eu/Fe]<1) or MP-$r$II( [Eu/Fe]>1) could form.  

\section{The evolutionary channels of the fast merging DNS systems}\label{ss.channels}

Approximately 10-20 Myr is required to evolve a high-mass binary from birth to the formation of a double neutron star system. The subsequent time to merger due to gravitational wave radiation varies by many orders of magnitude, but can be of order Myr or less. As indicated in the previous section, there are two general evolutionary scenarios for forming DNS with short merging times:

(i) During the second SN, a system can randomly receive a kick with the right magnitude and direction to place it on a highly eccentric orbit. 
This occurs in a small but non-negligible fraction of DNSs, and quickly leads to a merger ($\displaystyle{\lim_{e \to 1} t_{\rm merge} = 0}$). 
We call this the highly eccentric orbit channel and discuss it further in subsection \ref{ss.highE}.

(ii) During the late stages of evolution, a system can experience unstable mass transfer from a He-star donor to a NS accretor that does not lead to a merger within the CE. This evolutionary channel results in the formation of a binary system of a NS with a naked CO-core, and orbital separation $\approx0.1\rsun$. Due to their tight orbital separations, these systems likely survive the second SN regardless of the kick velocity and direction to form a DNS that quickly merges afterward. We call this the unstable case BB mass transfer channel and discuss it further in subsection \ref{ss.BB}.

\subsection{The Highly Eccentric Orbit Channel}\label{ss.highE}

After the formation of the first NS in a DNS, a typical high-mass binary will evolve through one or more mass transfer phases (systems that avoid mass transfer are too widely separated to merge within a Hubble time). 
The first of these phases is typically a CE \citep{MacLeod:2015bw},
forming a tight binary comprised of a NS with a He-star companion that has lost its H-envelope. After further evolution, these ``naked'' He-stars begin to form a stratified structure of a Carbon-Oxygen (CO) core, surrounded by a He-envelope, in the process evolving through a Helium HG phase. Such ``naked'' He HG stars typically have radii $\approx \rsun$. Depending on the orbital separation, the system may enter a second phase of mass transfer -- the so-called Case BB mass transfer phase -- in which the NS accretes He from the He HG star \citep{Delgado:1981vo}. Depending on the stars' masses at this stage and the binary's separation, this mass transfer may be either stable or unstable \citep{Ivanova:2003hc}, and we consider the stable case in this section \citep[For a detailed picture of this evolutionary channel, see][Figure 2]{Chruslinska:2018bv}.

From the population synthesis results, we find that systems going through this channel have separations $\approx \rsun$ immediately prior to the second SN. DNSs formed through this channel comprise the bulk of the mostly green distributions in the bottom panels of Figure \ref{f:all_models}. In order to obtain intuition for the impact of natal kicks on the resulting DNSs, we show the distribution of merger times (after the second SN) in the top panel of Figure \ref{f.kicks} for four different kick velocity distributions for a circular pre-SN orbit with an orbital separation of 1$\rsun$. These are produced using the analytic formulation described in Andrews \& Zezas (submitted) assuming NSs with masses of 1.4$\msun$ NSs and a 2.0$\msun$ pre-SN helium star mass. 

Because of the strong dependence of the merger time on the DNS orbital separation ($t_{\rm merge} \sim a^4$), the merger time of DNSs formed through this scenario peaks at $\approx$100 Myr. Figure \ref{f.kicks} indicates that merger times of $<$1 Myr are possible through this formation scenario, but only with relatively large kick velocities and then only at the tail of the distribution. These systems with short merger times are produced when ``lucky" kicks pointed in the optimal direction (opposite to the orbital velocity), produce DNS with small orbital separations and large eccentricities. Since this channel requires large NS kick velocities, our low kick velocity models for sub-model B produce fewer candidate DNSs than our high kick velocity models. However, regardless of the kick velocity applied, systems with merger times of $\approx$1 Myr are rare for systems formed through this evolutionary channel (for example $\approx1\%$ of all the DNSs in BZ002H). These fast merging binaries form the types of DNS that can merge within a UFD and are denoted as purple points in sub-model B. This channel exists in both sub-models A and B, however, as we will show, sub-model A provides yet another channel to produce our candidate DNSs. 

\subsection{The Unstable Case BB Mass Transfer Channel} \label{ss.BB}

Previous evolutionary simulations of mass transfer from a He HG star onto a NS \citep{Ivanova:2003hc,Dewi:2003dy} indicate that some systems, depending on the component masses and orbital period, should go through unstable Case BB mass transfer. However, the outcome of unstable Case BB mass transfer is uncertain; these systems enter into Roche Lobe Overflow as He HG stars, and as such, the donor stars in these systems lack clear core-envelope boundaries. For example, recent work by \cite{VignaGomez:2018th} suggests that this Case BB phase ought to be predominantly stable to reproduce the Galactic DNS population. 
Moreover, recent simulations suggest that this phase of mass transfer may be stable \citep{Tauris:2013,Lazarus:2014,Tauris:2015}.
Nevertheless, uncertainty in the evolution of this phase has led us to test both sub-model A, in which systems are allowed to survive unstable case BB mass transfer, and sub-model B, in which such systems are forced to merge. Therefore the formation of DNSs through this evolutionary channel \citep[For a detailed picture of this evolutionary channel, see Figure C1 in][ and references therein]{Chruslinska:2018bv}, only occurs in our sub-model A. The result of binaries in sub-model A that survive unstable case BB mass transfer is a tightly-bound DNS with a separation of $\approx$0.1$\rsun$.

\begin{figure}
\setlength{\tabcolsep}{1mm}
\hspace{0cm}
\includegraphics[width=1.0\columnwidth]{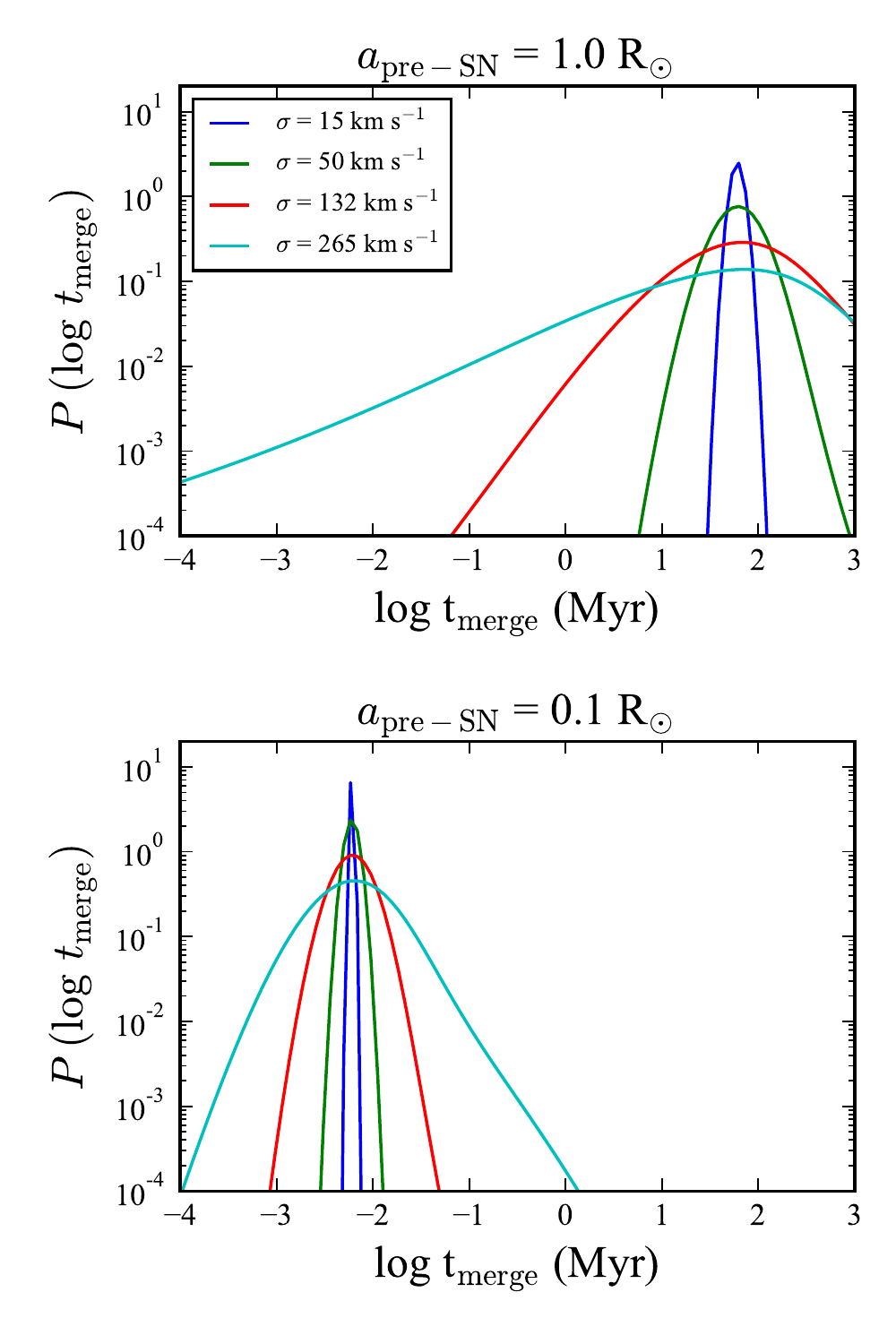}
\caption{The merger time distribution (taken to be the time between DNS formation and merger) for systems with orbital separations characteristic of formation through an evolutionary channel with either stable (top panel) or unstable (bottom panel) Case BB mass transfer. The curves are computed analytically. For small kicks, the distributions are sharply peaked, while as the kick distributions increase, the merger time distributions broaden substantially. Formation of short merger time DNSs through the standard evolutionary channel (top panel) is only possible for systems formed with a large kick velocity, and then only for a small minority of systems at the tail of the distribution. 
}
\label{f.kicks}
\end{figure}

The bottom panel of Figure \ref{f.kicks} shows the merger time distributions for DNSs with a pre-SN orbital separation of 0.1$\rsun$. For small kicks (relative to the orbital velocity), the NS natal kick only induces a relatively small perturbation on top of the orbit, leading to a sharply peaked merger time distribution at $\sim$10$^{-2}$ Myr. For larger kick velocity distributions, the merger time distribution broadens, as a larger range of DNS orbits become possible. Yet, despite the variation in the kick velocity distribution, the merger time is still strongly peaked between 10$^{-3}$ to 10$^{-1}$ Myr, and the corresponding fraction of candidate DNSs differs by less than a factor of two.

\section{Comparing to observations}\label{f.obs}
In this section, we compare the total stellar mass needed to form in order to make one DNS candidate in the stellar mass content of the UFDs. The stellar mass content of Reticulum II is about $\approx2.6^{+0.2}_{-0.2}\times10^3\msun$ \citep{Bechtol:2015bd}, and the current estimate for Tucana III is about $\approx$8$\times10^2$$\msun$ \citep{DrlicaWagner:2015gb} which makes it the lowest-luminosity ultra-faint dwarf found to contain a $r$-I star \citep{Hansen:2017ho}. 

Despite their low masses today, there is evidence that these systems could have initially been more massive \citep{Penarrubia:2008eq}. For instance, the observed tidal tails around Tucana III indicate that about 80\% of the stellar mass is in the process of being tidally stripped \citep{Shipp:2018tb,Li:2018vi}, which would make the total stellar mass of Tucana III  $\approx 4.6\times10^3\msun$. 

\begin{figure}
\setlength{\tabcolsep}{1mm}
\hspace{0cm}
\includegraphics[width=1.0\columnwidth]{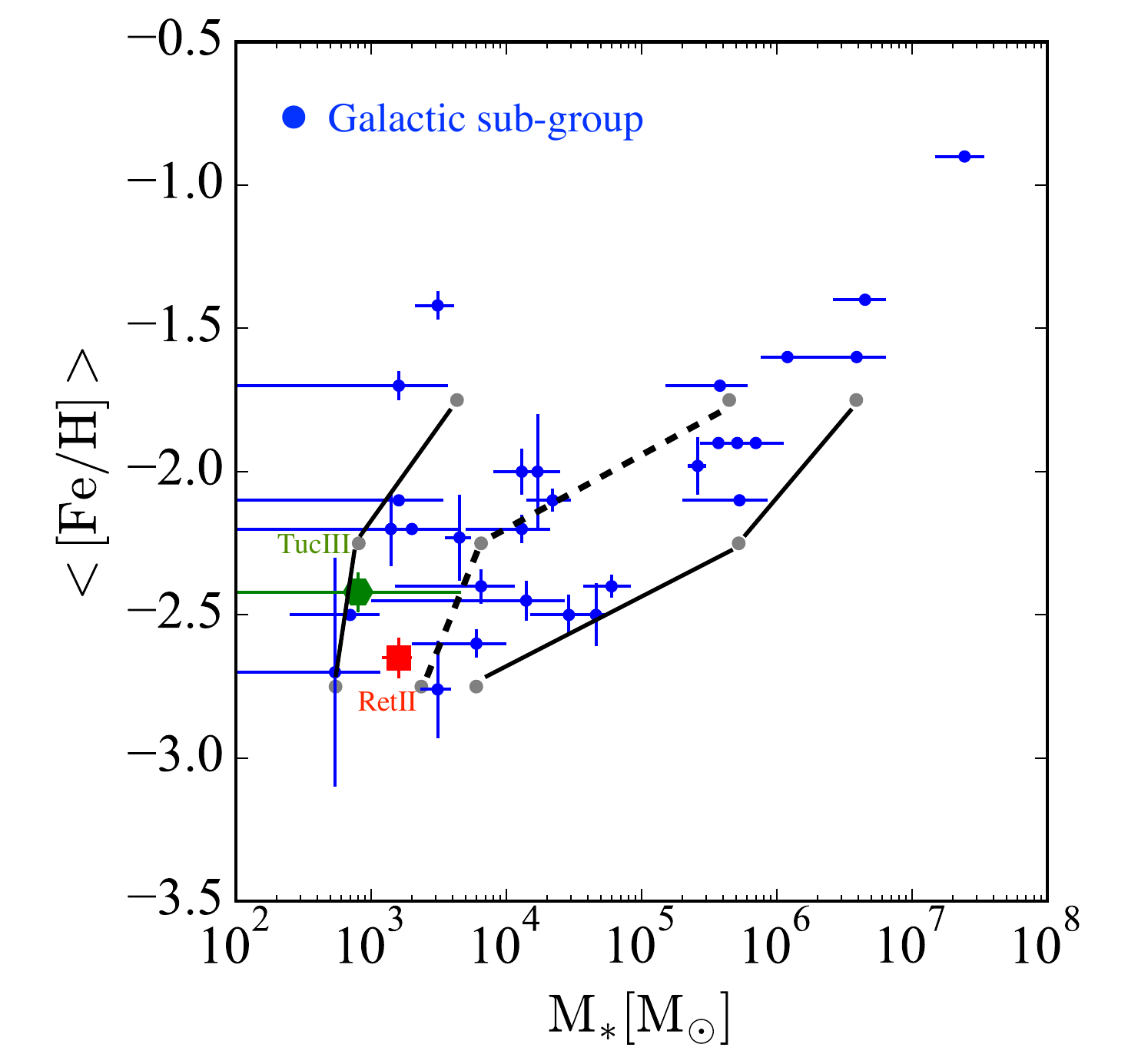}
\caption{Shows the stellar metallicity vs.\ stellar mass of the MW's satellites in solid blue circles adopted from the \citet{McConnachie:2012fh} compilation, with updates from \cite{Fattahi:2018cf}. Tucana III and Reticulum II are shown with red square and green hexagon, respectively, and their corresponding data are adopted 
from \citep{DrlicaWagner:2015gb,Simon:2017ij} and \citep{Bechtol:2015bd,Simon:2015ih}. The dashed line indicates the median value stellar mass for three different metallicity bins of $-3<[{\rm Fe/H}]<-2.5$, $-2.5<[{\rm Fe/H}]<-2$, and $-2<[{\rm Fe/H}]<-1.5$, while solid lines encompass the $3-\sigma$ envelope around the median. Horizontal distance of the galaxies from the right edge of the $3-\sigma$ envelope could potentially indicate how much stellar mass in these systems have been lost through tidal stripping, assuming tides do not change the metallicities \citep{Fattahi:2018cf}.}
\label{f.MvFeH}
\end{figure}

Figure \ref{f.MvFeH} shows the stellar metallicity vs.\ absolute visual magnitude of the MW's satellites \citep{McConnachie:2012fh}, where the V band absolute magnitude of Tucana III and Reticulum II are $-2.4\pm{0.42}$ \citep{DrlicaWagner:2015gb} and $-3.6\pm{0.1}$ \citep{Bechtol:2015bd}.
The corresponding metallicities of these two systems are $-2.42 \pm 0.07$ \citep{Simon:2017ij}, and $-2.65 \pm 0.07$ \citep{Simon:2015ih}. 
The flattening of the data toward lower luminosities has been attributed to tidal stripping \citep{Fattahi:2018cf}.

The horizontal distance of the galaxies from this line could potentially indicate how much mass has been lost through tidal stripping in these systems, as tides do not affect the metallicities. 
The distances of the Tucana III and Reticulum II from the edge of the $3-\sigma$ envelope could indicate the original stellar masses of these systems. 
Estimating the stellar mass loss due to tidal stripping by drawing a horizontal line in metallicity-stellar mass plane results in initial stellar masses of Reticulum II and Tucana III to be about a factor of 10 and 40 more massive in the past, respectively. Since these estimates are consistent with the location of these systems in the circular velocity, $\rm V_{1/2}$, measured at the stellar half-mass radius ($r_{1/2}$) vs. $r_{1/2}$ plane \citep{Fattahi:2018cf}, 
we use the factors of 10 and 40 for the initial stellar masses of Reticulum II and Tucana III, respectively.

\begin{figure*}
\setlength{\tabcolsep}{1mm}
\hspace{0cm}
\includegraphics[width=1\textwidth,height=14cm,keepaspectratio]{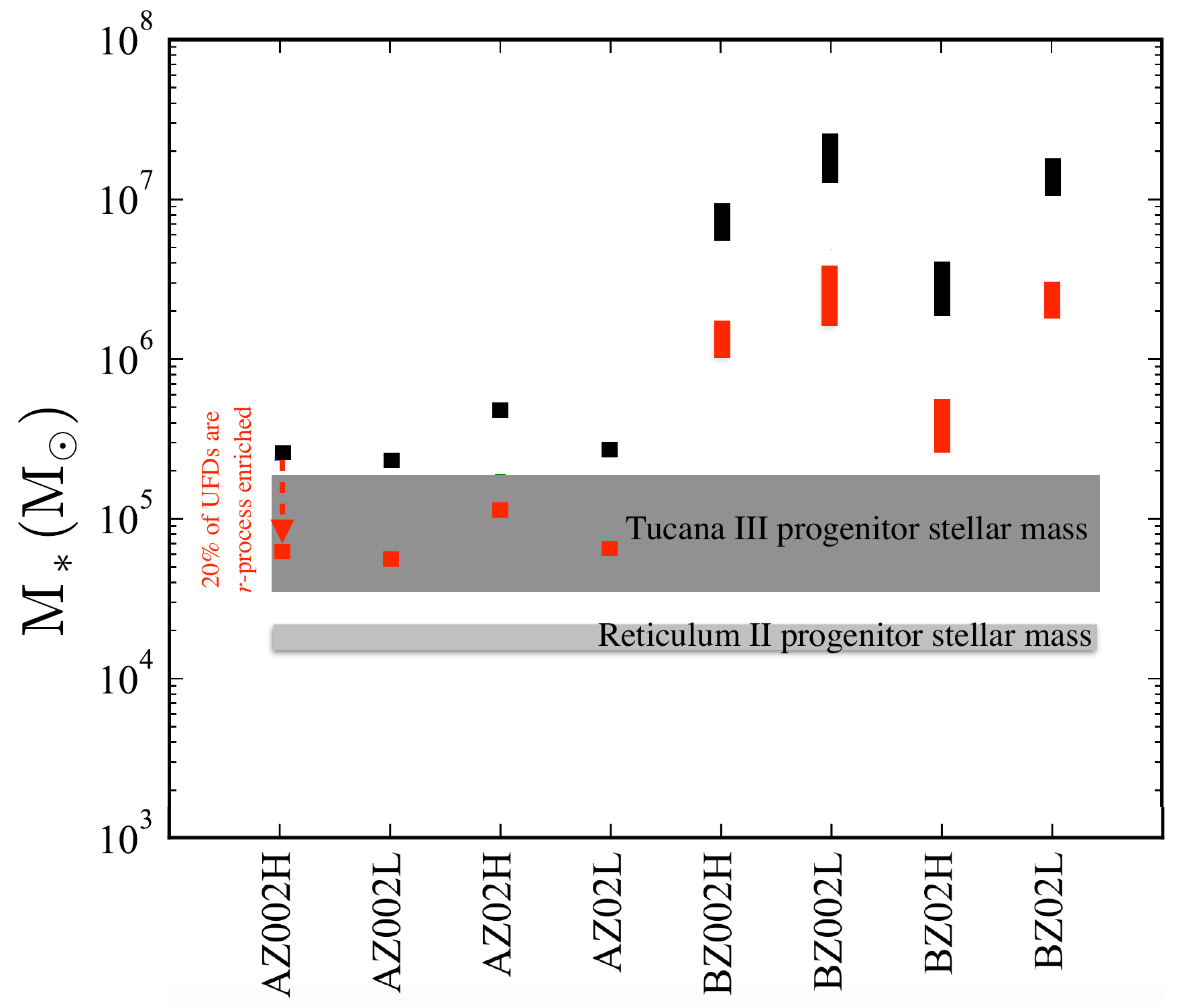}
\caption{The formation rate of candidate NS binaries. Each black bar indicates the total mass of stars needed to form in order to have one candidate DNS. 
There are 8 black lines corresponding to 8 different models studied here. The models are described in section 2.1. 
The length of each bar indicates the estimated birth rate of the candidates depending on the halo mass in which we assume the DNS is born in. We have considered two cases of $10^9\msun$ halo with $r_{\rm vir}\approx 4.5$ kpc)  at $z=6$, and $10^8\msun$ with $r_{\rm vir}\approx 1.3$ kpc at $z=10$ \citep{Safarzadeh:2018fi}.
The shaded regions indicate the estimated stellar mass of Reticulum II and Tucana III \citep{Bechtol:2015bd,DrlicaWagner:2015gb,Shipp:2018tb,Li:2018vi} after correcting for the tidal stripping mass loss. The downward thin red dashed arrow is of 0.7 dex size and indicates the fact that only about 20\%  of the total observed UFDs have been shown to be
enriched in $r$-process, and therefore, about an order of magnitude lower birth rates for the candidate DNSs are allowed. All the red bars are the black bars shifted downward by 0.7 dex.}
\label{f.summary}
\end{figure*}

Figure \ref{f.summary} summarizes our results.
Here the black bars indicate the total stellar mass needed to form in order to make one candidate DNS in each model. The height of each bar indicates different assumptions about the virial radius of a UFD progenitor halo. 

We have considered two cases of $10^9\msun$ halo with $r_{\rm vir}\approx 4.5$ kpc)  at $z=6$, and $10^8\msun$ halo with $r_{\rm vir}\approx 1.3$ kpc at $z=10$.  The required stellar mass in order to form a candidate DNS ranges from $10^5\msun$ for submodel A to $10^7$ for submodel B. Note that only a fraction (10-20\%) of UFDs
within 100 kpc of the the MW's center
are observed to be \rprocess enriched.

To reflect this, the black bars in Figure \ref{f.summary} are lowered by 0.7 dex (assuming only about 20\% of the UFDs are observed to be \rprocess enriched) and are shown as red bars. In other words, although about $3\times 10^5\msun$ of stellar mass is needed to form in order to make one DNS candidate in AZ002H model, a lower formation rate is possible since not all the UFDs are \rprocess enriched. 

The stellar masses of the two observed \rprocess UFDs are shown with shaded regions where the width of the shaded regions for each of the satellites indicates the lower and upper limits on their stellar mass estimates after correcting for the stellar mass loss due to tidal stripping.

As illustrated in Figure \ref{f.summary}, assuming these UFD systems have been tidally stripped, the predicted formation rate in submodel A, where HG star survive the common envelope phase, makes NSMs a plausible source of \rprocess enrichment in such systems. Although the estimated progenitor stellar mass of Reticulum II is  a factor few below the red bars, this is still consistent with the overall picture we provide in this paper, because the estimated total stellar mass that needs to form in order to make $2\times10^6$ binaries based on the prescription in \citet{Dominik:2012cw} can be lower than $2.8\times10^8\msun$ by slight modifications to the model. Moreover, if Tucana III is a globular cluster and not a UFD, the red bars would need to be 1 dex lower than the black bars which would make the results more consistent with Reticulum II.   

\section{Summary and Discussion}

NSMs produce copious amount of \rprocess  material, but it is difficult to reconcile the large natal kicks of NS binaries (DNSs) with the levels of \rprocess enrichment seen in ultra-faint dwarf (UFD) galaxies such as Reticulum II and Tucana III.
Here we have used a standard binary population synthesis model to identify a subset of DNSs whose
{\em combination} of systemic center of mass velocities and merging times
make them merge well within the virial radius of their host halo, before they travel far from the star-forming region of their host galaxy. 

The simulations have been carried out at different metallicity bins, different natal kick PDFs, and two different assumptions regarding unstable mass transfer with HG donors: Submodel A considers the separation of the core and envelope in HG donor stars while in submodel B this separation is not modeled, and therefore a CE phase in submodel B always leads to merging of the compact object with the secondary stars' core. Due to this reason, a smaller percentage of all the compact binaries that are formed, end up as NS binaries in submodel B relative to submodel A. 

We conclude that only submodel A produces a population of DNSs that can \rprocess enrich UFD-type galaxies, assuming, of course, that a large fraction of the \rprocess material ejected in an NSM event taking place well within the virial radius is recycled and will be locked in the stars formed after the event.  
In particular, one can form the candidate DNSs through a second phase of unstable mass transfer that shrinks the orbital separation to $\approx 0.1\rsun$ and produces DNS merger timescales of $<$Myr after the second SN. 
Although DNSs formed in highly eccentric orbits are considered as candidates, their formation rate is very low and therefore would not make for a population that can explain \rprocess UFDs by themselves. 
Binaries that shrink to small orbital separations through a second phase of mass transfer are similar to the ultra stripped helium stars \citep{Tauris:2017cf}.

These results, however, rely on the assumption that more than 90\% of the initial stellar mass of the observed \rprocess UFDs have been tidally stripped since entering the MW's halo progenitor. The observations of Tucana III show that at least 80\% of its stellar mass is being tidally stripped \citep{Shipp:2018tb,Li:2018vi}. Tidal stripping of satellites of a MW type halo has been studied extensively and have been shown that a large fraction of stellar mass is indeed stripped away \citep[e.g., see ][]{Penarrubia:2008eq}. However, the claim that these systems are tidally stripped, could be challenging to prove theoretically. \citet{vandenBosch:2018fv} show that due to adiabatic resilience, 
disruption of the subhalos in CDM paradigm is rare and even if the imparted energy to the subhalo exceeds its binding energy by a factor of 100, the subhalo still survives as a bound structure. Therefore, they
claim that in the absence of baryonic processes all the disruptions of the subhalos in CDM simulations should be a numerical artifact. Although there are observations that support the tidal stripping scenario \citep{Collins:2017ix}, Future WFIRST observations of the field UFD galaxies that are free from
tidal stripping scenarios can shed light on this matter.

On the other hand, recent deep HST/ACS observations of the UFDs hint to the presence of a top-heavy IMF in such systems \citep{Gennaro:2018bq}. This result is interesting in that 
\citet{deMink:2015bh} have shown a more top-heavy IMF will lead to a factor of 2 or 3 increase in the overall formation of the NS binaries. Although this by itself is not enough to explain 
the observed statistics of the \rprocess enhanced UFDs, it nevertheless helps to alleviates the tension.
Moreover, as mentioned in the previous section, we have assumed that the total stellar mass budget of $2\times10^6$ binaries could be tweaked within a factor of 2 assuming different IMFs. This point, could be rather crucial for the conclusions of this works because the red bars in Figure 5 stand above the stellar mass range of Ret-II.

Dynamical assembly in clusters at high redshifts has been proposed to produce tight DNSs \citep{RamirezRuiz:2015gl} with short merging timescales. However, the small masses of UFDs disfavor a dynamical assembly origin for the candidate DNSs. Fast merging channels in triple systems \citep{Bonetti:2018co} could boost the merging rates, but the models analyzed in this work do not consider this channel. 
Separately, magnetorotationally-driven (MRD) supernovae \citep{Winteler:2012fv,Wehmeyer:2015kl,Nishimura:2015fv,Mosta:2018ip,Halevi:2018gq} 
have been proposed to explain the observed high stellar abundances of Eu and Ba in such systems since
MRDs provide copious amount of \rprocess material. However, MRDs have been excluded from being a candidate since the evolution of [Mg/Fe] and [Ba/Fe] vs. [Fe/H] in dwarf galaxies suggest different sources for $\alpha$ elements and \rprocess material \citep{Duggan:2018wq}.

In this work, we have only considered models that are publicly available for analysis.
These either have solar or 0.1 solar metallicities, while UFDs have metallicities 
of $\rm [Fe/H]<-2$ which is 10 times lower than our lowest metallicity bin.
The extrapolation of our results to lower metallicities might appear as a stretch as 
many aspects of stellar evolution are metallicity dependent in a non-linear way (e.g., stellar winds).
However, DNS production efficiency either remains unchanged at lower metallicities or increase: \citet{Bramante:2016kp} have shown that NS production efficiency is relatively 
unchanged when comparing the simulation with $\rm [Fe/H]=-2.3$ to those with $\rm [Fe/H]=-1$, and \citet{Chruslinska:2018bv} show that the merger rate of the DNSs increase in their standard model with decreasing the metallicities from 0.1 solar to 0.01 solar (see their table 2), which is in favor of our proposed model.  

In the end, we need more complete models of fast merging DNSs as \rprocess sources in UFDs. 
These would come from performing high resolution zoom cosmological simulations on UFD host halos while modeling the candidate DNSs as \rprocess sources with their
characteristic natal kicks and merging times.
While we have shown in \citet{Safarzadeh:2017dq} that a single NSM event can explain Reticulum II like systems, these sets of simulations are vital to perform, since otherwise
the impact of natal kicks remains unclear. The ejected \rprocess material might not get effectively recycled into the new generation of the stars \citep{Macias:2018br}. This is because \rprocess material will be preferentially deposited in the halo and might get ejected before getting recycled.
These simulations will predict the level of \rprocess enhancement, i.e., the ratio of different class of \rprocess enhanced stars to their parent category, that would be expected to observe in an ensemble of 
\rprocess UFDs if fast merging NSMs are considered as the source of \rprocess enrichment in the UFDs. 

\acknowledgements
 We are thankful to the referee for their careful reading of our work and helpful comments. We would like to thank Selma de Mink, Massimo Ricotti, Marla Geha, Frank van den Bosch, and David Radice for useful discussions. M.T.S. and J.J.A. are thankful to Niels Bohr Institute for hospitality which made this work possible. J.J.A. acknowledges funding from the European Research Council under the European Union's Seventh Framework Program (FP/2007-2013)/ERC Grant Agreement n.\ 617001. M.T.S. and E.S. were supported by NASA theory grant NNX15AK82G. ER thanks the DNRF for support as a Niels Bohr Professor.

\bibliographystyle{apj}
\bibliography{the_entire_lib}

\begin{thebibliography}{}
\expandafter\ifx\csname natexlab\endcsname\relax\def\natexlab#1{#1}\fi

\bibitem[{Abbott {et~al.}(2017)Abbott, Abbott, Abbott, Acernese, Ackley, Adams,
  Adams, Addesso, Adhikari, Adya, Affeldt, Afrough, Agarwal, Agathos, Agatsuma,
  Aggarwal, Aguiar, Aiello, Ain, Ajith, Allen, Allen, Allocca, Altin, Amato,
  Ananyeva, Anderson, Anderson, Angelova, Antier, Appert, Arai, Araya, Areeda,
  Arnaud, Arun, Ascenzi, Ashton, Ast, Aston, Astone, Atallah, Aufmuth, Aulbert,
  AultONeal, Austin, Avila-Alvarez, Babak, Bacon, Bader, Bae, Baker,
  Baldaccini, Ballardin, Ballmer, Banagiri, Barayoga, Barclay, Barish, Barker,
  Barkett, Barone, Barr, Barsotti, Barsuglia, Barta, Barthelmy, Bartlett,
  Bartos, Bassiri, Basti, Batch, Bawaj, Bayley, Bazzan, B{\'e}csy, Beer,
  Bejger, Belahcene, Bell, Berger, Bergmann, Bero, Berry, Bersanetti,
  Bertolini, Betzwieser, Bhagwat, Bhandare, Bilenko, Billingsley, Billman,
  Birch, Birney, Birnholtz, Biscans, Biscoveanu, Bisht, Bitossi, Biwer,
  Bizouard, Blackburn, Blackman, Blair, Blair, Blair, Bloemen, Bock, Bode,
  Boer, Bogaert, Bohe, Bondu, Bonilla, Bonnand, Boom, Bork, Boschi, Bose,
  Bossie, Bouffanais, Bozzi, Bradaschia, Brady, Branchesi, Brau, Briant,
  Brillet, Brinkmann, Brisson, Brockill, Broida, Brooks, Brown, Brown, Brunett,
  Buchanan, Buikema, Bulik, Bulten, Buonanno, Buskulic, Buy, Byer, Cabero,
  Cadonati, Cagnoli, Cahillane, Bustillo, Callister, Calloni, Camp, Canepa,
  Canizares, Cannon, Cao, Cao, Capano, Capocasa, Carbognani, Caride, Carney,
  Diaz, Casentini, Caudill, Cavagli{\`a}, Cavalier, Cavalieri, Cella, Cepeda,
  Cerd{\'a}-Dur{\'a}n, Cerretani, Cesarini, Chamberlin, Chan, Chao, Charlton,
  Chase, Chassande-Mottin, Chatterjee, Chatziioannou, Cheeseboro, Chen, Chen,
  Chen, Cheng, Chia, Chincarini, Chiummo, Chmiel, Cho, Cho, Chow, Christensen,
  Chu, Chua, Chua, Chung, Chung, Ciani, Ciolfi, Cirelli, Cirone, Clara, Clark,
  Clearwater, Cleva, Cocchieri, Coccia, Cohadon, Cohen, Colla, Collette,
  Cominsky, Jr, Conti, Cooper, Corban, Corbitt, Cordero-Carri{\'o}n, Corley,
  Cornish, Corsi, Cortese, Costa, Coughlin, Coughlin, Coulon, Countryman,
  Couvares, Covas, Cowan, Coward, Cowart, Coyne, Coyne, Creighton, Creighton,
  Cripe, Crowder, Cullen, Cumming, Cunningham, Cuoco, Canton, D{\'a}lya,
  Danilishin, D'Antonio, Danzmann, Dasgupta, Da~Silva~Costa, Dattilo, Dave,
  Davier, Davis, Daw, Day, De, DeBra, Degallaix, Laurentis, Del{\'e}glise,
  Pozzo, Demos, Denker, Dent, Pietri, Dergachev, Rosa, DeRosa, Rossi, DeSalvo,
  Varona, Devenson, Dhurandhar, D{\'\i}az, Fiore, Giovanni, Girolamo, Lieto,
  Pace, Palma, Renzo, Doctor, Dolique, Donovan, Dooley, Doravari, Dorrington,
  Douglas, {\'A}lvarez, Downes, Drago, Dreissigacker, Driggers, Du, Ducrot,
  Dupej, Dwyer, Edo, Edwards, Effler, Ehrens, Eichholz, Eikenberry, \&
  Eisenste...}]{Abbott:2017it}
Abbott, B.~P., Abbott, R., Abbott, T.~D., {et~al.} 2017, The Astrophysical
  Journal, 848, L12

\bibitem[{Abbott et~al .(2017)}]{Collaboration:2017kt}
Abbott et~al ., B.~P. 2017, arXiv.org, 161101

\bibitem[{Andrews {et~al.}(2015)Andrews, Farr, Kalogera, \&
  Willems}]{Andrews:2015bq}
Andrews, J.~J., Farr, W.~M., Kalogera, V., \& Willems, B. 2015, The
  Astrophysical Journal, 801, 32

\bibitem[{Bechtol {et~al.}(2015)Bechtol, Drlica-Wagner, Balbinot, Pieres,
  Simon, Yanny, Santiago, Wechsler, Frieman, Walker, Williams, Rozo, Rykoff,
  Queiroz, Luque, Benoit-Levy, Tucker, Sevilla, Gruendl, da~Costa, Fausti~Neto,
  Maia, Abbott, Allam, Armstrong, Bauer, Bernstein, Bernstein, Bertin, Brooks,
  Buckley-Geer, Burke, Carnero~Rosell, Castander, Covarrubias, D'Andrea, DePoy,
  Desai, Diehl, Eifler, Estrada, Evrard, Fernandez, Finley, Flaugher,
  Gaztanaga, Gerdes, Girardi, Gladders, Gruen, Gutierrez, Hao, Honscheid, Jain,
  James, Kent, Kron, Kuehn, Kuropatkin, Lahav, Li, Lin, Makler, March,
  Marshall, Martini, Merritt, Miller, Miquel, Mohr, Neilsen, Nichol, Nord,
  Ogando, Peoples, Petravick, Plazas, Romer, Roodman, Sako, Sanchez, Scarpine,
  Schubnell, Smith, Soares-Santos, Sobreira, Suchyta, Swanson, Tarle, Thaler,
  Thomas, Wester, Zuntz, \& des Collaboration}]{Bechtol:2015bd}
Bechtol, K., Drlica-Wagner, A., Balbinot, E., {et~al.} 2015, The Astrophysical
  Journal, 807, 50

\bibitem[{Beers \& Christlieb(2005)}]{Beers:2005kn}
Beers, T.~C., \& Christlieb, N. 2005, Annual Review of Astronomy and
  Astrophysics, 43, 531

\bibitem[{Behroozi {et~al.}(2014)Behroozi, Ramirez-Ruiz, \&
  Fryer}]{Behroozi:2014bp}
Behroozi, P.~S., Ramirez-Ruiz, E., \& Fryer, C.~L. 2014, The Astrophysical
  Journal, 792, 123

\bibitem[{Belczynski {et~al.}(2010)Belczynski, Bulik, Fryer, Ruiter, Valsecchi,
  Vink, \& Hurley}]{Belczynski:2010iw}
Belczynski, K., Bulik, T., Fryer, C.~L., {et~al.} 2010, The Astrophysical
  Journal, 714, 1217

\bibitem[{Belczynski {et~al.}(2002)Belczynski, Kalogera, \&
  Bulik}]{Belczynski:2002gi}
Belczynski, K., Kalogera, V., \& Bulik, T. 2002, The Astrophysical Journal,
  572, 407

\bibitem[{Belczynski {et~al.}(2008)Belczynski, Kalogera, Rasio, Taam, Zezas,
  Bulik, Maccarone, \& Ivanova}]{Belczynski:2008kt}
Belczynski, K., Kalogera, V., Rasio, F.~A., {et~al.} 2008, The Astrophysical
  Journal Supplement Series, 174, 223

\bibitem[{Belczynski {et~al.}(2006)Belczynski, Perna, Bulik, Kalogera, Ivanova,
  \& Lamb}]{Belczynski:2006dq}
Belczynski, K., Perna, R., Bulik, T., {et~al.} 2006, The Astrophysical Journal,
  648, 1110

\bibitem[{{Belczynski} {et~al.}(2007){Belczynski}, {Taam}, {Kalogera}, {Rasio},
  \& {Bulik}}]{Belczynski:2007}
{Belczynski}, K., {Taam}, R.~E., {Kalogera}, V., {Rasio}, F.~A., \& {Bulik}, T.
  2007, \apj, 662, 504

\bibitem[{Beniamini {et~al.}(2016)Beniamini, Hotokezaka, \&
  Piran}]{Beniamini:2016kwa}
Beniamini, P., Hotokezaka, K., \& Piran, T. 2016, The Astrophysical Journal
  Letters, 829, L13

\bibitem[{Bland-Hawthorn {et~al.}(2015)Bland-Hawthorn, Sutherland, \&
  Webster}]{BlandHawthorn:2015ke}
Bland-Hawthorn, J., Sutherland, R., \& Webster, D. 2015, The Astrophysical
  Journal, 807, 154

\bibitem[{Bonetti {et~al.}(2018)Bonetti, Perego, Capelo, Dotti, \&
  Miller}]{Bonetti:2018co}
Bonetti, M., Perego, A., Capelo, P.~R., Dotti, M., \& Miller, M.~C. 2018,
  Publications of the Astronomical Society of Australia, 35, 124021

\bibitem[{Bovill \& Ricotti(2009)}]{Bovill:2009hg}
Bovill, M.~S., \& Ricotti, M. 2009, The Astrophysical Journal, 693, 1859

\bibitem[{Bovill \& Ricotti(2011)}]{Bovill:2011bk}
---. 2011, The Astrophysical Journal, 741, 17

\bibitem[{Bramante \& Linden(2016)}]{Bramante:2016kp}
Bramante, J., \& Linden, T. 2016, The Astrophysical Journal, 826, 57

\bibitem[{Brown {et~al.}(2012)Brown, Tumlinson, Geha, Kirby, VandenBerg,
  Mu{\~n}oz, Kalirai, Simon, Avila, Guhathakurta, Renzini, \&
  Ferguson}]{Brown:2012jo}
Brown, T.~M., Tumlinson, J., Geha, M., {et~al.} 2012, The Astrophysical Journal
  Letters, 753, L21

\bibitem[{Brown {et~al.}(2014)Brown, Tumlinson, Geha, Simon, Vargas,
  VandenBerg, Kirby, Kalirai, Avila, Gennaro, Ferguson, Mu{\~n}oz,
  Guhathakurta, \& Renzini}]{Brown:2014jn}
---. 2014, The Astrophysical Journal, 796, 91

\bibitem[{Chruslinska {et~al.}(2018)Chruslinska, Belczynski, Klencki, \&
  Benacquista}]{Chruslinska:2018bv}
Chruslinska, M., Belczynski, K., Klencki, J., \& Benacquista, M. 2018, Monthly
  Notices of the Royal Astronomical Society, 474, 2937

\bibitem[{Collins {et~al.}(2017)Collins, Tollerud, Sand, Bonaca, Willman, \&
  Strader}]{Collins:2017ix}
Collins, M. L.~M., Tollerud, E.~J., Sand, D.~J., {et~al.} 2017, Monthly Notices
  of the Royal Astronomical Society, stx067

\bibitem[{Coulter {et~al.}(2017)Coulter, Foley, Kilpatrick, Drout, Piro,
  Shappee, Siebert, Simon, Ulloa, Kasen, Madore, Murguia-Berthier, Pan,
  Prochaska, Ramirez-Ruiz, Rest, \& Rojas-Bravo}]{Coulter:2017ei}
Coulter, D.~A., Foley, R.~J., Kilpatrick, C.~D., {et~al.} 2017, Science, 358,
  1556

\bibitem[{de~Mink \& Belczynski(2015)}]{deMink:2015bh}
de~Mink, S.~E., \& Belczynski, K. 2015, The Astrophysical Journal, 814, 58

\bibitem[{Delgado \& Thomas(1981)}]{Delgado:1981vo}
Delgado, A.~J., \& Thomas, H.~C. 1981, Astronomy {\&} Astrophysics, 96, 142

\bibitem[{Deloye \& Taam(2010)}]{Deloye:2010ey}
Deloye, C.~J., \& Taam, R.~E. 2010, The Astrophysical Journal, 719, L28

\bibitem[{Dewi \& Pols(2003)}]{Dewi:2003dy}
Dewi, J. D.~M., \& Pols, O.~R. 2003, Monthly Notices of the Royal Astronomical
  Society, 344, 629

\bibitem[{Dominik {et~al.}(2012)Dominik, Belczynski, Fryer, Holz, Berti, Bulik,
  Mandel, \& O'Shaughnessy}]{Dominik:2012cw}
Dominik, M., Belczynski, K., Fryer, C., {et~al.} 2012, The Astrophysical
  Journal, 759, 52

\bibitem[{Drlica-Wagner {et~al.}(2015)Drlica-Wagner, Bechtol, Rykoff, Luque,
  Queiroz, Mao, Wechsler, Simon, Santiago, Yanny, Balbinot, Dodelson,
  Fausti~Neto, James, Li, Maia, Marshall, Pieres, Stringer, Walker, Abbott,
  Abdalla, Allam, Benoit-Levy, Bernstein, Bertin, Brooks, Buckley-Geer, Burke,
  Carnero~Rosell, Carrasco~Kind, Carretero, Crocce, da~Costa, Desai, Diehl,
  Dietrich, Doel, Eifler, Evrard, Finley, Flaugher, Fosalba, Frieman,
  Gaztanaga, Gerdes, Gruen, Gruendl, Gutierrez, Honscheid, Kuehn, Kuropatkin,
  Lahav, Martini, Miquel, Nord, Ogando, Plazas, Reil, Roodman, Sako, Sanchez,
  Scarpine, Schubnell, Sevilla-Noarbe, Smith, Soares-Santos, Sobreira, Suchyta,
  Swanson, Tarle, Tucker, Vikram, Wester, Zhang, Zuntz, \& des
  Collaboration}]{DrlicaWagner:2015gb}
Drlica-Wagner, A., Bechtol, K., Rykoff, E.~S., {et~al.} 2015, The Astrophysical
  Journal, 813, 109

\bibitem[{{Drout} {et~al.}(2017){Drout}, {Piro}, {Shappee}, {Kilpatrick},
  {Simon}, {Contreras}, {Coulter}, {Foley}, {Siebert}, {Morrell}, {Boutsia},
  {Di Mille}, {Holoien}, {Kasen}, {Kollmeier}, {Madore}, {Monson},
  {Murguia-Berthier}, {Pan}, {Prochaska}, {Ramirez-Ruiz}, {Rest}, {Adams},
  {Alatalo}, {Ba{\~n}ados}, {Baughman}, {Beers}, {Bernstein}, {Bitsakis},
  {Campillay}, {Hansen}, {Higgs}, {Ji}, {Maravelias}, {Marshall}, {Moni Bidin},
  {Prieto}, {Rasmussen}, {Rojas-Bravo}, {Strom}, {Ulloa},
  {Vargas-Gonz{\'a}lez}, {Wan}, \& {Whitten}}]{Drout:2017gb}
{Drout}, M.~R., {Piro}, A.~L., {Shappee}, B.~J., {et~al.} 2017, Science, 358,
  1570

\bibitem[{Duggan {et~al.}(2018)Duggan, Kirby, Andrievsky, \&
  Korotin}]{Duggan:2018wq}
Duggan, G.~E., Kirby, E.~N., Andrievsky, S.~M., \& Korotin, S.~A. 2018,
  1809.04597

\bibitem[{Fattahi {et~al.}(2018)Fattahi, Navarro, Frenk, Oman, Sawala, \&
  Schaller}]{Fattahi:2018cf}
Fattahi, A., Navarro, J.~F., Frenk, C.~S., {et~al.} 2018, Monthly Notices of
  the Royal Astronomical Society, 476, 3816

\bibitem[{Frebel \& Bromm(2012)}]{Frebel:2012ja}
Frebel, A., \& Bromm, V. 2012, The Astrophysical Journal, 759, 115

\bibitem[{Fryer {et~al.}(2012)Fryer, Belczynski, Wiktorowicz, Dominik,
  Kalogera, \& Holz}]{Fryer:2012jk}
Fryer, C.~L., Belczynski, K., Wiktorowicz, G., {et~al.} 2012, The Astrophysical
  Journal, 749, 91

\bibitem[{Gennaro {et~al.}(2018)Gennaro, Tchernyshyov, Brown, Geha, Avila,
  Guhathakurta, Kalirai, Kirby, Renzini, Simon, Tumlinson, \&
  Vargas}]{Gennaro:2018bq}
Gennaro, M., Tchernyshyov, K., Brown, T.~M., {et~al.} 2018, The Astrophysical
  Journal, 855, 20

\bibitem[{{Giacobbo} \& {Mapelli}(2018)}]{Giacobbo:2018}
{Giacobbo}, N., \& {Mapelli}, M. 2018, \mnras, 480, 2011

\bibitem[{Halevi \& M{\"o}sta(2018)}]{Halevi:2018gq}
Halevi, G., \& M{\"o}sta, P. 2018, Monthly Notices of the Royal Astronomical
  Society, 477, 2366

\bibitem[{Hansen {et~al.}(2017)Hansen, Simon, Marshall, Li, Carollo, DePoy,
  Nagasawa, Bernstein, Drlica-Wagner, Abdalla, Allam, Annis, Bechtol,
  Benoit-Levy, Brooks, Buckley-Geer, Carnero~Rosell, Carrasco~Kind, Carretero,
  Cunha, da~Costa, Desai, Eifler, Fausti~Neto, Flaugher, Frieman,
  Garc{\'\i}a-Bellido, Gaztanaga, Gerdes, Gruen, Gruendl, Gschwend, Gutierrez,
  James, Krause, Kuehn, Kuropatkin, Lahav, Miquel, Plazas, Romer, Sanchez,
  Santiago, Scarpine, Smith, Soares-Santos, Sobreira, Suchyta, Swanson, Tarle,
  Walker, \& des Collaboration}]{Hansen:2017ho}
Hansen, T.~T., Simon, J.~D., Marshall, J.~L., {et~al.} 2017, The Astrophysical
  Journal, 838, 44

\bibitem[{Hobbs {et~al.}(2005)Hobbs, Lorimer, Lyne, \& Kramer}]{Hobbs:2005be}
Hobbs, G., Lorimer, D.~R., Lyne, A.~G., \& Kramer, M. 2005, Monthly Notices of
  the Royal Astronomical Society, 360, 974

\bibitem[{Ivanova {et~al.}(2003)Ivanova, Belczynski, Kalogera, Rasio, \&
  Taam}]{Ivanova:2003hc}
Ivanova, N., Belczynski, K., Kalogera, V., Rasio, F.~A., \& Taam, R.~E. 2003,
  The Astrophysical Journal, 592, 475

\bibitem[{{Janka}(2017)}]{Janka:2017}
{Janka}, H.-T. 2017, \apj, 837, 84

\bibitem[{Ji {et~al.}(2016)Ji, Frebel, Chiti, \& Simon}]{Ji:2016ja}
Ji, A.~P., Frebel, A., Chiti, A., \& Simon, J.~D. 2016, Nature, 531, 610

\bibitem[{Ji {et~al.}(2018)Ji, Simon, Frebel, Venn, \& Hansen}]{Ji:2018wr}
Ji, A.~P., Simon, J.~D., Frebel, A., Venn, K.~A., \& Hansen, T.~T. 2018, eprint
  arXiv:1809.02182, 1809.02182

\bibitem[{Kasen {et~al.}(2017)Kasen, Metzger, Barnes, Quataert, \&
  Ramirez-Ruiz}]{Kasen:2017kk}
Kasen, D., Metzger, B., Barnes, J., Quataert, E., \& Ramirez-Ruiz, E. 2017,
  Nature, 551, 80

\bibitem[{Koposov {et~al.}(2015{\natexlab{a}})Koposov, Belokurov, Torrealba, \&
  Evans}]{Koposov:2015cw}
Koposov, S.~E., Belokurov, V., Torrealba, G., \& Evans, N.~W.
  2015{\natexlab{a}}, The Astrophysical Journal, 805, 130

\bibitem[{Koposov {et~al.}(2015{\natexlab{b}})Koposov, Casey, Belokurov, Lewis,
  Gilmore, Worley, Hourihane, Randich, Bensby, Bragaglia, Bergemann, Carraro,
  Costado, Flaccomio, Francois, Heiter, Hill, Jofre, Lando, Lanzafame,
  de~Laverny, Monaco, Morbidelli, Sbordone, Mikolaitis, \&
  Ryde}]{Koposov:2015ep}
Koposov, S.~E., Casey, A.~R., Belokurov, V., {et~al.} 2015{\natexlab{b}}, The
  Astrophysical Journal, 811, 62

\bibitem[{Kravtsov(2013)}]{Kravtsov:2013cy}
Kravtsov, A.~V. 2013, The Astrophysical Journal Letters, 764, L31

\bibitem[{{Lazarus} {et~al.}(2014){Lazarus}, {Tauris}, {Knispel}, {Freire},
  {Deneva}, {Kaspi}, {Allen}, {Bogdanov}, {Chatterjee}, {Stairs}, \&
  {Zhu}}]{Lazarus:2014}
{Lazarus}, P., {Tauris}, T.~M., {Knispel}, B., {et~al.} 2014, \mnras, 437, 1485

\bibitem[{Li {et~al.}(2018)Li, Simon, Kuehn, Pace, Erkal, Bechtol, Yanny,
  Drlica-Wagner, Marshall, Lidman, Balbinot, Carollo, Jenkins,
  Martinez-Vazquez, Shipp, Stringer, Vivas, Walker, Wechsler, Abdalla, Allam,
  Annis, Avila, Bertin, Brooks, Buckley-Geer, Burke, Rosell, Kind, Carretero,
  Cunha, D'Andrea, da~Costa, Davis, De~Vicente, Doel, Eifler, Evrard, Flaugher,
  Frieman, Garc{\'\i}a-Bellido, Gaztanaga, Gerdes, Gruen, Gruendl, Gschwend,
  Gutierrez, Hartley, Hollowood, Honscheid, James, Krause, Maia, March,
  Menanteau, Miquel, Plazas, Sanchez, Santiago, Scarpine, Schindler, Schubnell,
  Sevilla-Noarbe, Smith, Smith, Soares-Santos, Sobreira, Suchyta, Swanson,
  Tarle, \& Tucker}]{Li:2018vi}
Li, T.~S., Simon, J.~D., Kuehn, K., {et~al.} 2018, 1804.07761

\bibitem[{Macias \& Ramirez-Ruiz(2018)}]{Macias:2018br}
Macias, P., \& Ramirez-Ruiz, E. 2018, The Astrophysical Journal, 860, 89

\bibitem[{MacLeod \& Ramirez-Ruiz(2015)}]{MacLeod:2015bw}
MacLeod, M., \& Ramirez-Ruiz, E. 2015, The Astrophysical Journal, 798, L19

\bibitem[{{Margutti} {et~al.}(2018){Margutti}, {Alexander}, {Xie}, {Sironi},
  {Metzger}, {Kathirgamaraju}, {Fong}, {Blanchard}, {Berger}, {MacFadyen},
  {Giannios}, {Guidorzi}, {Hajela}, {Chornock}, {Cowperthwaite}, {Eftekhari},
  {Nicholl}, {Villar}, {Williams}, \& {Zrake}}]{Margutti:2018ek}
{Margutti}, R., {Alexander}, K.~D., {Xie}, X., {et~al.} 2018, \apjl, 856, L18

\bibitem[{McConnachie(2012)}]{McConnachie:2012fh}
McConnachie, A.~W. 2012, The Astronomical Journal, 144, 4

\bibitem[{M{\"o}sta {et~al.}(2018)M{\"o}sta, Roberts, Halevi, Ott, Lippuner,
  Haas, \& Schnetter}]{Mosta:2018ip}
M{\"o}sta, P., Roberts, L.~F., Halevi, G., {et~al.} 2018, The Astrophysical
  Journal, 864, 171

\bibitem[{Nishimura {et~al.}(2015)Nishimura, Takiwaki, \&
  Thielemann}]{Nishimura:2015fv}
Nishimura, N., Takiwaki, T., \& Thielemann, F.-K. 2015, The Astrophysical
  Journal, 810, 109

\bibitem[{Penarrubia {et~al.}(2008)Penarrubia, Navarro, \&
  McConnachie}]{Penarrubia:2008eq}
Penarrubia, J., Navarro, J.~F., \& McConnachie, A.~W. 2008, The Astrophysical
  Journal, 673, 226

\bibitem[{{Pfahl} {et~al.}(2002){Pfahl}, {Rappaport}, {Podsiadlowski}, \&
  {Spruit}}]{Pfahl:2002}
{Pfahl}, E., {Rappaport}, S., {Podsiadlowski}, P., \& {Spruit}, H. 2002, \apj,
  574, 364

\bibitem[{Ramirez-Ruiz {et~al.}(2015)Ramirez-Ruiz, Trenti, MacLeod, Roberts,
  Lee, \& Saladino-Rosas}]{RamirezRuiz:2015gl}
Ramirez-Ruiz, E., Trenti, M., MacLeod, M., {et~al.} 2015, The Astrophysical
  Journal Letters, 802, L22

\bibitem[{Safarzadeh \& C{\^o}t{\'e}(2017)}]{Safarzadeh:2017dw}
Safarzadeh, M., \& C{\^o}t{\'e}, B. 2017, Monthly Notices of the Royal
  Astronomical Society, 471, 4488

\bibitem[{Safarzadeh {et~al.}(2018)Safarzadeh, Ji, Dooley, Frebel, Scannapieco,
  G{\'o}mez, \& O'Shea}]{Safarzadeh:2018fi}
Safarzadeh, M., Ji, A.~P., Dooley, G.~A., {et~al.} 2018, Monthly Notices of the
  Royal Astronomical Society, 476, 5006

\bibitem[{Safarzadeh \& Scannapieco(2017)}]{Safarzadeh:2017dq}
Safarzadeh, M., \& Scannapieco, E. 2017, Monthly Notices of the Royal
  Astronomical Society, 471, 2088

\bibitem[{Shipp {et~al.}(2018)Shipp, Drlica-Wagner, Balbinot, Ferguson, Erkal,
  Li, Bechtol, Belokurov, Buncher, Carollo, Carrasco~Kind, Kuehn, Marshall,
  Pace, Rykoff, Sevilla-Noarbe, Sheldon, Strigari, Vivas, Yanny, Zenteno,
  Abbott, Abdalla, Allam, Avila, Bertin, Brooks, Burke, Carretero, Castander,
  Cawthon, Crocce, Cunha, D'Andrea, da~Costa, Davis, De~Vicente, Desai, Diehl,
  Doel, Evrard, Flaugher, Fosalba, Frieman, Garc{\'\i}a-Bellido, Gaztanaga,
  Gerdes, Gruen, Gruendl, Gschwend, Gutierrez, Hoyle, James, Johnson, Krause,
  Kuropatkin, Lahav, Lin, Maia, March, Martini, Menanteau, Miller, Miquel,
  Nichol, Plazas, Romer, Sako, Sanchez, Scarpine, Schindler, Schubnell, Smith,
  Smith, Sobreira, Suchyta, Swanson, Tarle, Thomas, Tucker, Walker, Wechsler,
  \& des Collaboration}]{Shipp:2018tb}
Shipp, N., Drlica-Wagner, A., Balbinot, E., {et~al.} 2018, eprint
  arXiv:1801.03097, 1801.03097

\bibitem[{Simon \& Geha(2007)}]{Simon:2007ee}
Simon, J.~D., \& Geha, M. 2007, The Astrophysical Journal, 670, 313

\bibitem[{Simon {et~al.}(2015)Simon, Drlica-Wagner, Li, Nord, Geha, Bechtol,
  Balbinot, Buckley-Geer, Lin, Marshall, Santiago, Strigari, Wang, Wechsler,
  Yanny, Abbott, Bauer, Bernstein, Bertin, Brooks, Burke, Capozzi, Rosell,
  Kind, D'Andrea, da~Costa, DePoy, Desai, Diehl, Dodelson, Cunha, Estrada,
  Evrard, Neto, Fernandez, Finley, Flaugher, Frieman, Gaztanaga, Gerdes, Gruen,
  Gruendl, Honscheid, James, Kent, Kuehn, Kuropatkin, Lahav, Maia, March,
  Martini, Miller, Miquel, Ogando, Romer, Roodman, Rykoff, Sako, Sanchez,
  Schubnell, Sevilla, Smith, Soares-Santos, Sobreira, Suchyta, Swanson, Tarle,
  Thaler, Tucker, Vikram, Walker, \& Wester}]{Simon:2015ih}
Simon, J.~D., Drlica-Wagner, A., Li, T.~S., {et~al.} 2015, arXiv.org, 95

\bibitem[{Simon {et~al.}(2017)Simon, Li, Drlica-Wagner, Bechtol, Marshall,
  James, Wang, Strigari, Balbinot, Kuehn, Walker, Abbott, Allam, Annis,
  Benoit-Levy, Brooks, Buckley-Geer, Burke, Carnero~Rosell, Carrasco~Kind,
  Carretero, Cunha, D'Andrea, da~Costa, DePoy, Desai, Doel, Fernandez,
  Flaugher, Frieman, Garc{\'\i}a-Bellido, Gaztanaga, Goldstein, Gruen,
  Gutierrez, Kuropatkin, Maia, Martini, Menanteau, Miller, Miquel, Neilsen,
  Nord, Ogando, Plazas, Romer, Rykoff, Sanchez, Santiago, Scarpine, Schubnell,
  Sevilla-Noarbe, Smith, Sobreira, Suchyta, Swanson, Tarle, Whiteway, Yanny, \&
  des Collaboration}]{Simon:2017ij}
Simon, J.~D., Li, T.~S., Drlica-Wagner, A., {et~al.} 2017, The Astrophysical
  Journal, 838, 11

\bibitem[{{Tauris} {et~al.}(2013){Tauris}, {Langer}, {Moriya}, {Podsiadlowski},
  {Yoon}, \& {Blinnikov}}]{Tauris:2013}
{Tauris}, T.~M., {Langer}, N., {Moriya}, T.~J., {et~al.} 2013, \apjl, 778, L23

\bibitem[{{Tauris} {et~al.}(2015){Tauris}, {Langer}, \&
  {Podsiadlowski}}]{Tauris:2015}
{Tauris}, T.~M., {Langer}, N., \& {Podsiadlowski}, P. 2015, \mnras, 451, 2123

\bibitem[{Tauris {et~al.}(2017)Tauris, Kramer, {Freire, P. C. C.}, Wex, Janka,
  Langer, Podsiadlowski, Bozzo, Chaty, Kruckow, van~den Heuvel, Antoniadis,
  Breton, \& Champion}]{Tauris:2017cf}
Tauris, T.~M., Kramer, M., {Freire, P. C. C.}, {et~al.} 2017, The Astrophysical
  Journal, 846, 170

\bibitem[{van~den Bosch {et~al.}(2018)van~den Bosch, Ogiya, Hahn, \&
  Burkert}]{vandenBosch:2018fv}
van~den Bosch, F.~C., Ogiya, G., Hahn, O., \& Burkert, A. 2018, Monthly Notices
  of the Royal Astronomical Society, 474, 3043

\bibitem[{{van den Heuvel}(2007)}]{van_den_Heuvel:2007}
{van den Heuvel}, E.~P.~J. 2007, in American Institute of Physics Conference
  Series, Vol. 924, The Multicolored Landscape of Compact Objects and Their
  Explosive Origins, ed. T.~{di Salvo}, G.~L. {Israel}, L.~{Piersant},
  L.~{Burderi}, G.~{Matt}, A.~{Tornambe}, \& M.~T. {Menna}, 598--606

\bibitem[{Vargas {et~al.}(2013)Vargas, Geha, Kirby, \& Simon}]{Vargas:2013ei}
Vargas, L.~C., Geha, M., Kirby, E.~N., \& Simon, J.~D. 2013, The Astrophysical
  Journal, 767, 134

\bibitem[{{Verbunt} {et~al.}(2017){Verbunt}, {Igoshev}, \&
  {Cator}}]{Verbunt:2017}
{Verbunt}, F., {Igoshev}, A., \& {Cator}, E. 2017, \aap, 608, A57

\bibitem[{Vigna-G{\'o}mez {et~al.}(2018)Vigna-G{\'o}mez, Neijssel, Stevenson,
  Barrett, Belczynski, Justham, de~Mink, M{\"u}ller, Podsiadlowski, Renzo,
  Sz{\'e}csi, \& Mandel}]{VignaGomez:2018th}
Vigna-G{\'o}mez, A., Neijssel, C.~J., Stevenson, S., {et~al.} 2018, eprint
  arXiv:1805.07974, 1805.07974

\bibitem[{Wehmeyer {et~al.}(2015)Wehmeyer, Pignatari, \&
  Thielemann}]{Wehmeyer:2015kl}
Wehmeyer, B., Pignatari, M., \& Thielemann, F.~K. 2015, Monthly Notices of the
  Royal Astronomical Society, 452, 1970

\bibitem[{Weisz {et~al.}(2014)Weisz, Dolphin, Skillman, Holtzman, Gilbert,
  Dalcanton, \& Williams}]{Weisz:2014cp}
Weisz, D.~R., Dolphin, A.~E., Skillman, E.~D., {et~al.} 2014, The Astrophysical
  Journal, 789, 148

\bibitem[{Winteler {et~al.}(2012)Winteler, K{\"a}ppeli, Perego, Arcones,
  Vasset, Nishimura, Liebend{\"o}rfer, \& Thielemann}]{Winteler:2012fv}
Winteler, C., K{\"a}ppeli, R., Perego, A., {et~al.} 2012, The Astrophysical
  Journal Letters, 750, L22

\end{thebibliography}

\end{document}